# SPECIALIZED DAY TRADING - A NEW VIEW ON AN OLD GAME

**Vladimir Šimović[1], and Vladimir Šimović[2], PhD**

[1] Faculty of Electrical Engineering and Computing, Unska 3, 10000 Zagreb, Croatia & Department of Mathematics, University of Zagreb, Bijenička cesta 30, 10000 Zagreb, Croatia, vladimir.simovic@fer.hr

[2] University of Zagreb, Faculty of Teacher Education, Savska cesta 77, Zagreb HR-10000, Croatia, vladimir.simovic@zg.htnet.hr

**Abstract**

After the U.S market earned strong returns in 2003, day trading made a comeback and once again became a „popular" trading method among traders. Although there is no comprehensive empirical evidence available to answer the question do individual day traders make money, there is a number of studies that point out that only few are able to consistently earn profits sufficient to cover transaction costs and thus make money. The day trading concept of buying and selling stocks on margin alone suggests that it is more risky than the usual "going long" way of making profit. This paper offers a new approach to day trading, an approach that eliminates some of the risks of day trading through specialization. The concept is that the trader should specialize himself in just one (blue chip) stock and use existing day trading techniques (trend following, playing news, range trading, scalping, technical analysis, covering spreads…) to make money.

## 1. Introduction

The name, day trading, refers to a practice of buying (selling short) and selling (buying to cover) stocks during the day in such manner, that at the end of the day there has been no net change in position; a complete round – trip trade has been made. A primary motivation of this style of trading is to avoid the risks of radical changes in prices that may occur if a stock is held overnight that could lead to large losses. Traders performing such round – trip trades are called day traders. The U.S. Securities and Exchange Commission adopted a new term in the year 2000, "pattern day trader", referring to a customer who places four or more round-trip orders over a five-day period, provided the number of trades is more than six percent in the account for the five day period.[1]

There are four common techniques used by day traders: trend following, playing news, range trading and scalping. Playing news and trend following are two techniques that are primarily in the realm of a day trader. When a trader is following a trend, he assumes that the stock which had been rising will continue to rise, and vice versa. One could say he is actually following the stocks

---

[1] On February 27, 2001, the Securities and Exchange Commission (SEC) approved amendments to National Association of Securities Dealers, Inc. (NASD®) Rule 2520 relating to margin requirements for day traders. Under the approved amendments, a pattern day trader would be required to maintain a minimum equity of $25,000 at all times. If the account falls below the $25,000 requirement, the pattern day trader would not be permitted to day trade until the account is restored.

"momentum". When a trader is playing news, his basic strategy is to buy a stock which has just announced good news, or sell short a stock which has announced bad news.

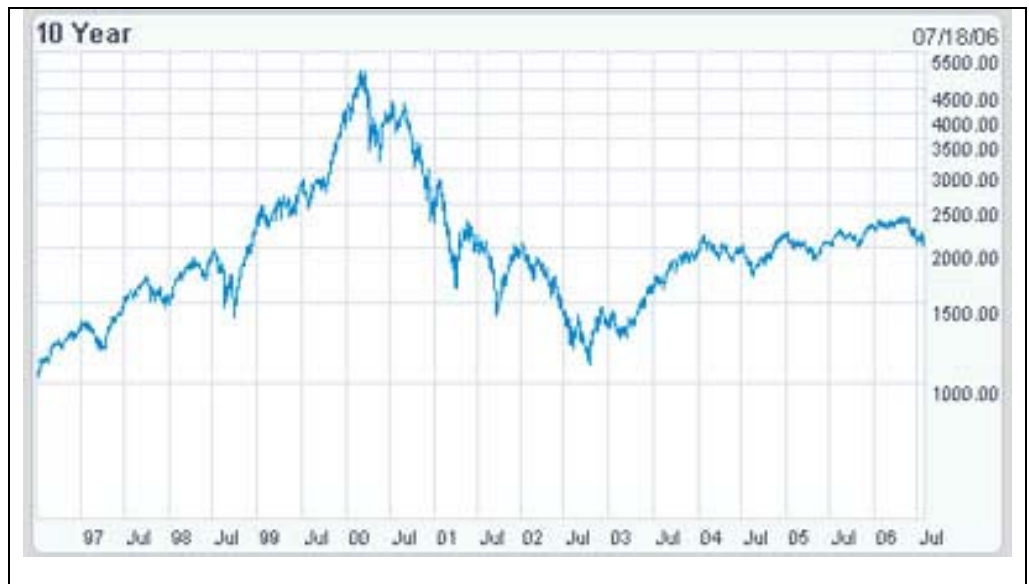

Figure 1 – NASDAQ[2] Composite quote data from 1997 to 2006

After its boom during the dotcom frenzy of the late 1990s and the loss in popularity after the Internet bubble burst, day trading is making a comeback. After three years of strong stock market performance, a constantly increasing number of investors use day trading techniques to make profit. A search on the Social Science Service Network reports 395 articles on day trading, with over 40% published in the last 3 years. A similar search on the most popular online bookstore, Amazon[3] will result in more than 400 popular books on day trading. Many of the popular news agencies and papers[4] report a surge in day trading popularity while some are also reporting its negative sides. Searching the most popular World Wide Web searching engine, Google[5], for the term "day trading" results in over 120,000,000 links. The first one of those links redirects a user's browser to a warning about risks involved in day trading published on the homepage of the U.S. Securities and Exchange Commission.

---

[2] National Association of Securities Dealers Automated Quotations
[3] According to Alexa Traffic Rankings, Amazon is the 15.th most popular site and the highest ranked online bookstore on the Top 500 site ranking list
[4] Associated Press reported a centrepiece "In Japan, day trading surges in popularity" on May 10, 2006. The Sunday Times published an article "High-risk day trading makes a comeback" on February 26, 2006
[5] Google is the most popular search engine according to the last Nielsen NetRatings Search Engine Ratings that were published in November of 2005

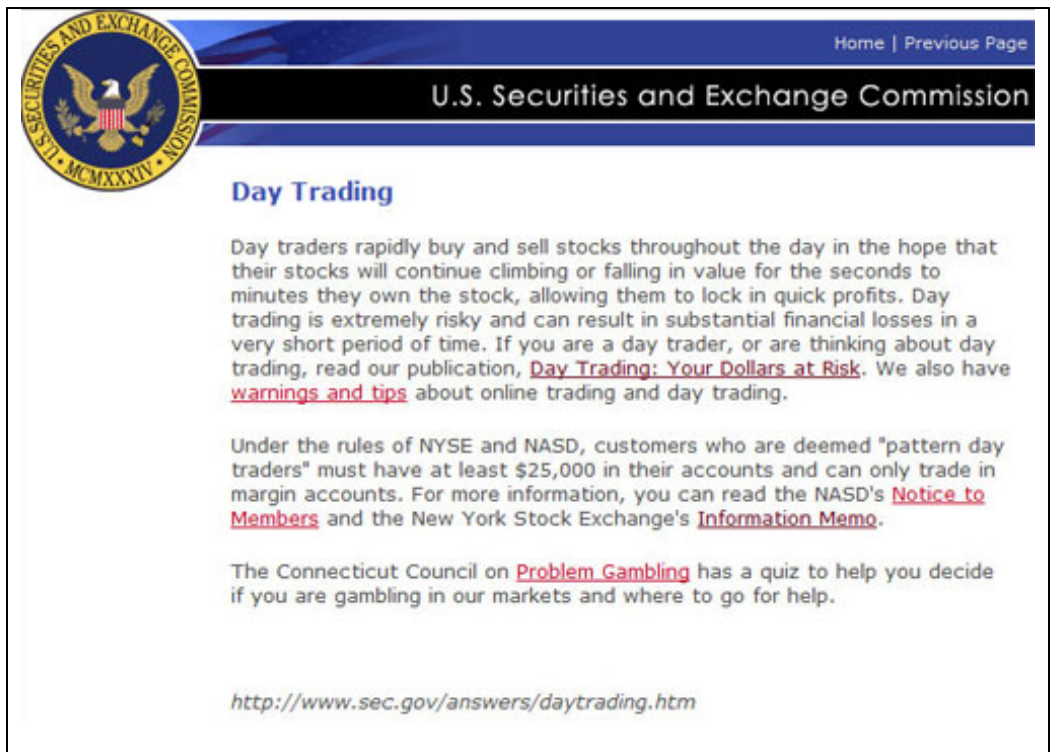

Figure 2 - U.S. Securities and Exchange Commission warning on day trading

## 2. The Controversy

The day trading controversy is mainly fuelled by its main con, it is risky. The constant usage of margin[6] is the strong and the weak point of day trading, because the usage of margin amplifies gains and losses such that substantial losses (and gains) may occur in a short period of time. Because day trading implies a minimum of two trades per business day[7], a part of the day trader's funds are used to pay commissions[8]. The higher the number of trades per day is, the bigger the part of day trader's funds is used to pay commissions. Day trading also often requires live quotes which are costly, and therefore also have an impact on the funds of a day trader. For every one of these (main) cons, day trading is, as it was already mentioned, considered risky.

An integral part in the day trading controversy is the day trader himself. Claims of easy and fast profits from day trading have attracted a significant number of non experienced and "casual" traders into day trading that do not fully understand the risks they are taking. With its latest comeback, day trading has become a business to people other than traders. Numerous websites offer tips and advices while online bookstores offer books on day trading strategies.

---

[6] borrowed funds
[7] buying (selling short) and selling (buying to cover)
[8] the broker's basic fee for purchasing or selling securities as an agent

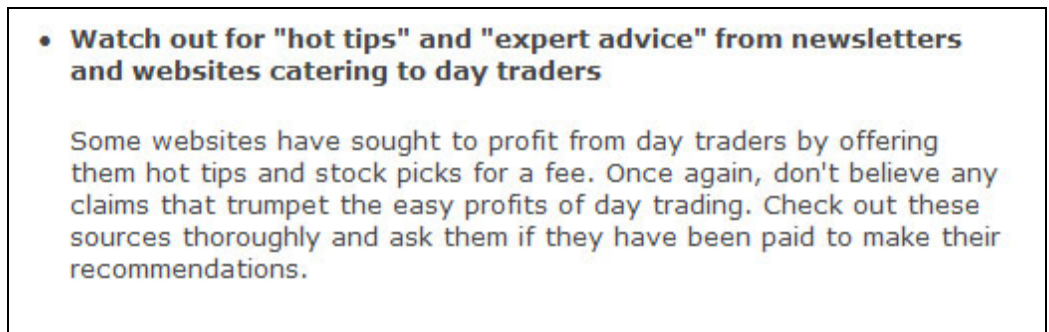

Figure 3 - U.S. Securities and Exchange Commission warning on day trading info

With all that in mind, one could wonder do day traders make money. Although that question can not be answered with certainty, a few existing studies do not paint a pretty picture. A comprehensive analysis of the profitability of all day trading activity in Taiwan over a five year period[9] has shown that in a typical six month period, more than eight out of ten day traders loose money.

## 3. Specialized Day Trading

### 3.1 The Concept

The main goal of Specialized Day Trading is to offer a new approach to day trading, an approach that would dampen or eliminate some of the negative sides of "regular" day trading. The concept is simple. Instead of using the usual day trading techniques on various types of stocks, the trader should specialize himself in using the same, already mentioned techniques, but with just one (blue chip[10]) stock.

### 3.2 Specialization

The main reason why specialization is proposed is in the fact that trading with different stocks on a daily basis brings a certain element of uncertainty since the day trader often does not have the time to thoroughly "check up" on a stock he is trading with. Focusing on just one stock eliminates the element of uncertainty and gives the day trader the opportunity, to through time better learn about its "behaviour" and how the selected stock reacts to certain events like splits, earning announcements, general (good or bad) news etc. Through constant monitoring, a trader could gain knowledge on how the stock reacts on markets ups and downs, better insight on the meaning of after hours trading activity or the manner how the company releases announcements (do worse announcements get announced on Friday to dampen the short-term response of the trader and thus the market?[11]) etc.

---

[9] Do Day Traders Make Money? Brad M. Barber, Yi-Tsung Lee, Yu-Jane Liu, Terrance Odean (January 2005)
[10] A blue chip stock is the stock of a well-established company having stable earnings and no extensive liabilities.
[11] Strategic Release of Information on Friday: Evidence from Earnings Announcements, Stefano Della Vigna, Joshua Pollet (January 2005)

### 3.3 Blue Chip

The focus was put on blue chip stocks because they offer stability, which of course translates into low risk. With day trading blue chip stocks one can not expect great profit in a single day, but the trade-off is that one can not expect great loss also. For further analysis of "behaviour" of the blue chip stocks, we've taken Microsoft as an example.

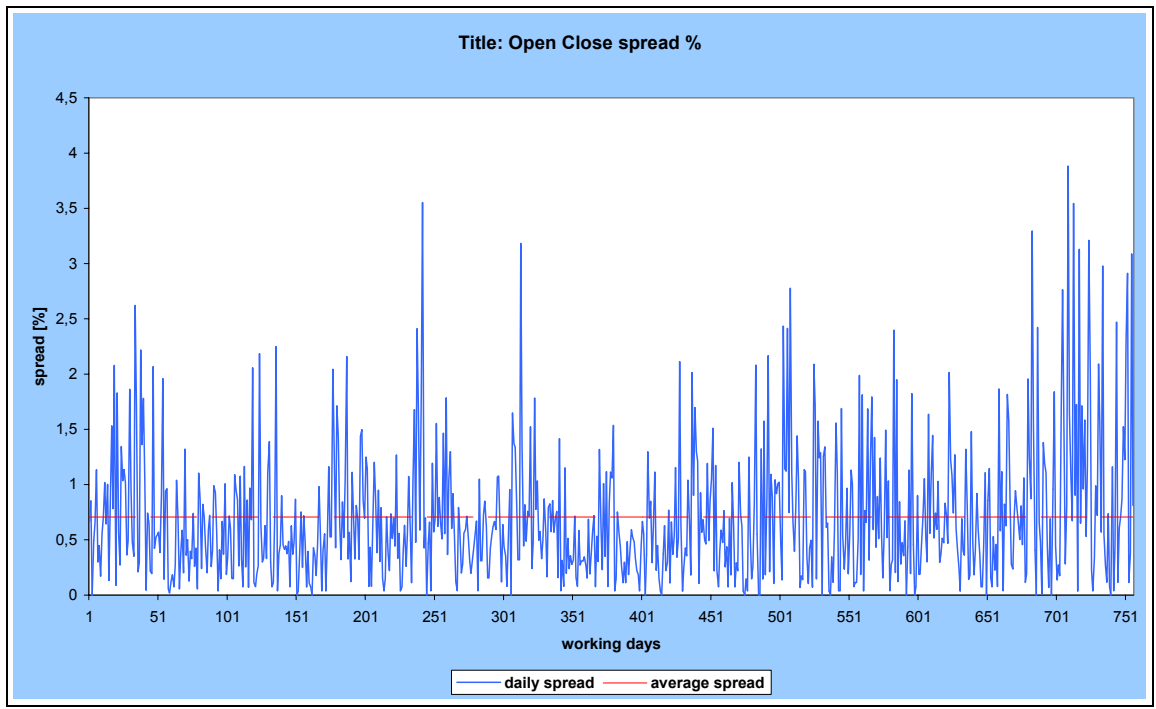

Figure 4 – a graph showing the spread between the opening and the closing price

The analysis was based on 3 years (756 working days) of daily data[12] (from 07/18/2003 till 07/18/2006). That particular period of time was chosen for our analysis because it consists of newest daily quotes from the last MSFT[13] stock split (02/18/2003). The data is consisted of the opening price ($o_i$), closing price ($c_i$), daily low ($l_i$) and the daily high ($h_i$). Using the following equations we've calculated the values (percentage) of the daily spreads ($S_i$) between the opening and the closing price. This was done in order to calculate the average spread ($S_{av}$) between the two already mentioned values which will help us illustrate the potential of this kind of day trading.

---

[12] The data was obtained on Nasdaq's website www.nasdaq.com
[13] Symbol that represents a Microsoft stock

$$\left| \frac{c_i - o_i}{o_i} * 100 \right| = S_i \qquad \frac{\sum_{i=1}^{n} S_i}{n} = S_{av} \qquad \begin{array}{l} i \in [1, n] \\ n = 756 \end{array}$$

The average spread between the opening and the closing price ($S_{av}$) for a Microsoft stock in that period of time was 0,706059344 % which is a very good indicator of its stability.

Using the exact equations we've also calculated the values (percentage) of the daily spreads ($Q_i$) between the daily low and the daily high. This was done in order to calculate the average spread ($Q_{av}$) between those two values so that we can illustrate the full (but unreachable) potential of this kind of day trading.

$$\left| \frac{h_i - l_i}{l_i} * 100 \right| = Q_i \qquad \frac{\sum_{i=1}^{n} Q_i}{n} = Q_{av} \qquad \begin{array}{l} i \in [1, n] \\ n = 756 \end{array}$$

The average spread between the daily low and the daily high ($Q_{av}$) for a Microsoft stock in that period of time was 1,57655549 %.

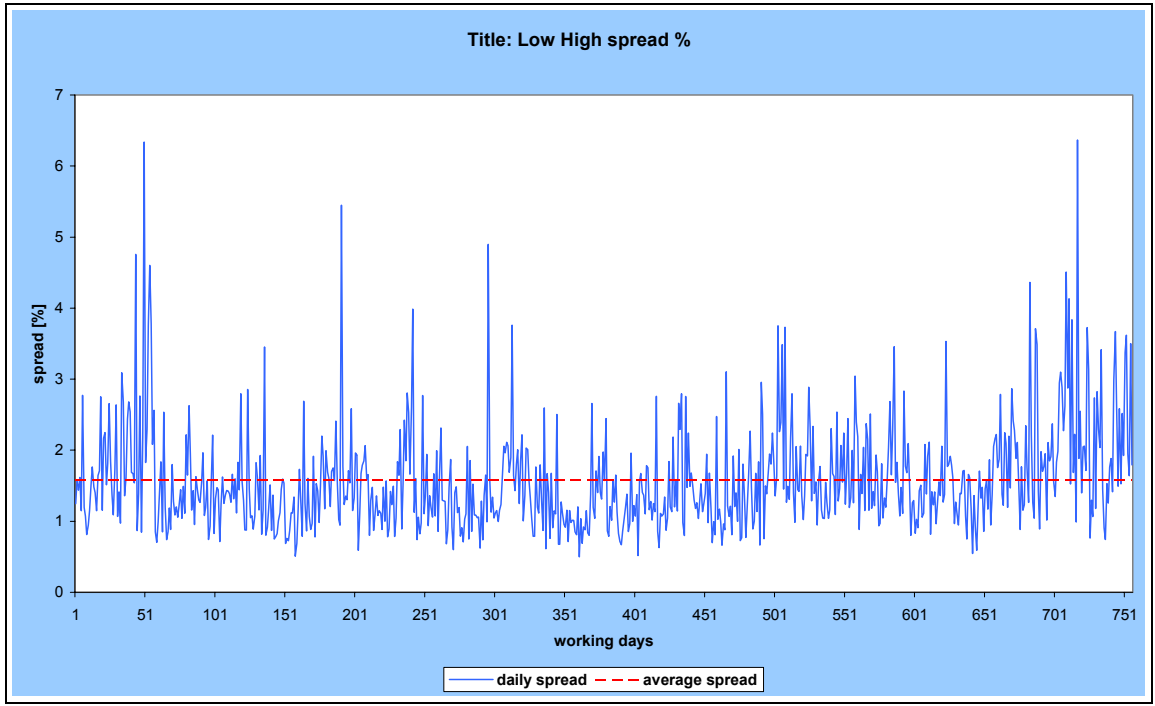

Figure 5 - a graph showing the spread between the daily low and the daily high

## 2.4 The Potential

To show the potential profit for this way of day trading we are going to use the following functions with the already calculated average daily spreads. These functions represent the total return (via percentage) in 30 working days of which a certain number (α) had a "positive" (and the rest "negative") trading outcome.

$$\left(1+\frac{S_{av}}{100}\right)^{\alpha} * \left(1-\frac{S_{av}}{100}\right)^{30-\alpha} * 100 = S_p$$

$$\left(1+\frac{Q_{av}}{100}\right)^{\alpha} * \left(1-\frac{Q_{av}}{100}\right)^{30-\alpha} * 100 = Q_p$$

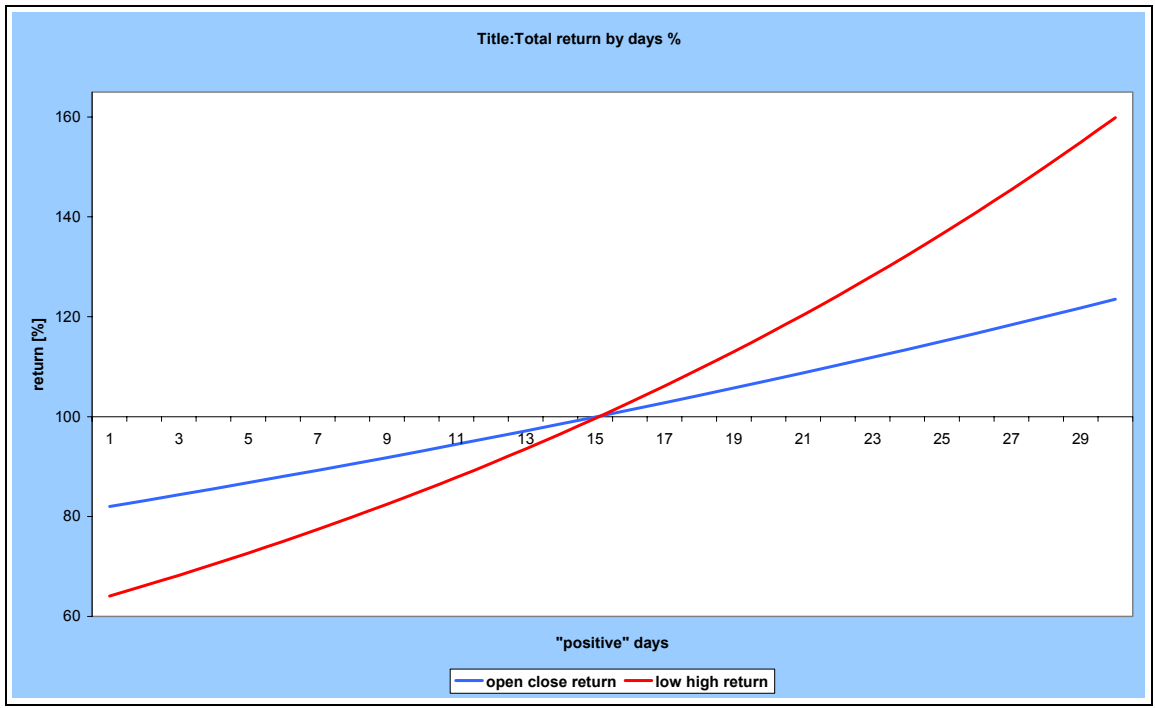

Figure 6 - a graph showing the spread between the daily low and the daily high without the usage of margin

With the usage of margin (M), gains and losses are amplified.

$$\left(1+\frac{S_{av}}{100}*\left(\frac{M}{100}\right)\right)^{\alpha} * \left(1-\frac{S_{av}}{100}*\left(\frac{M}{100}\right)\right)^{30-\alpha} * 100 = S_{mp}$$

$$\left(1+\frac{Q_{av}}{100}*\left(\frac{M}{100}\right)\right)^{\alpha} *\left(1-\frac{Q_{av}}{100}*\left(\frac{M}{100}\right)\right)^{30-\alpha} *100 = Q_{mp}$$

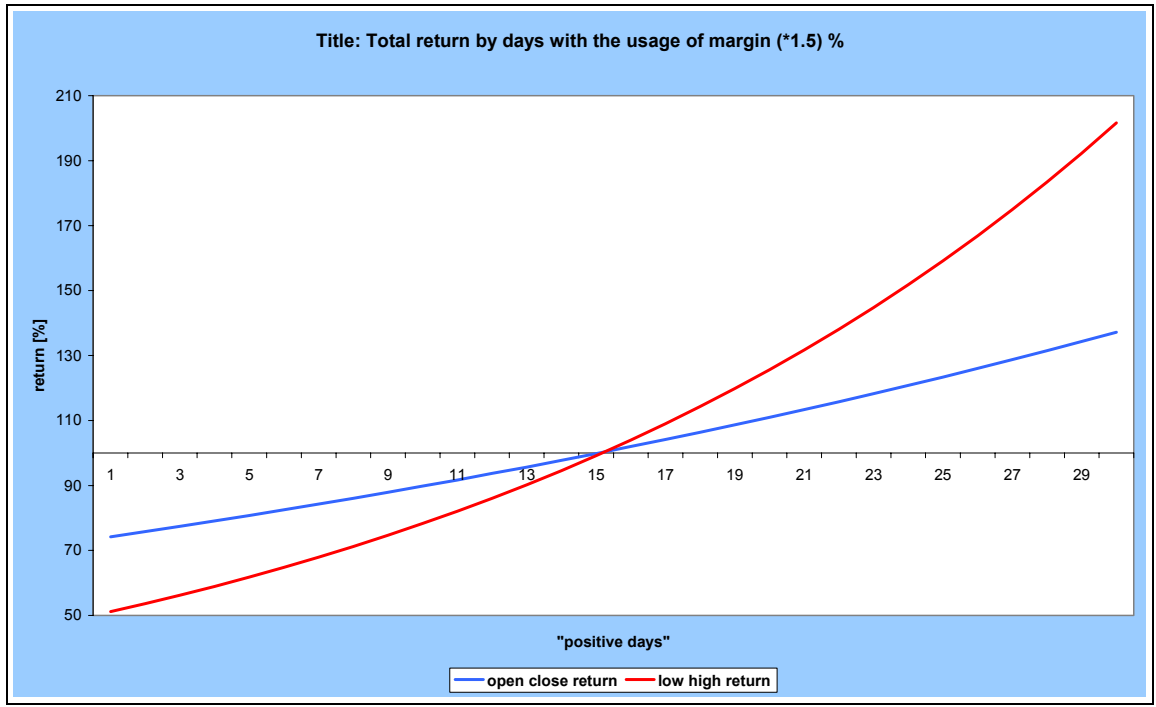

Figure 6 - a graph showing the spread between the daily low and the daily high with the usage of margin

## 4. Conclusion

Through Specialized Day Trading, we've tried to offers a new approach to day trading and with it eliminate some of the risks of day trading. The paper tried to explain the reasons behind the concept of specialization in trading in just one (blue chip) stock with the usage of existing day trading techniques and show that the usage of such concept has potential and can be profitable.